\documentclass[11pt]{article}
\usepackage[final]{acl}
\usepackage{times}
\usepackage{latexsym}
\usepackage[T1]{fontenc}
\usepackage[utf8]{inputenc}
\usepackage{microtype}
\usepackage{inconsolata}
\usepackage{booktabs,colortbl,multicol,multirow}
\usepackage{amsfonts,amsmath,amssymb}
\usepackage{arydshln}
\newcommand{\dval}[2]{\makebox[3.2em][r]{$#1$}\makebox[1.8em][l]{\,#2}}
\newcommand{\dvalci}[4]{\makebox[3.6em][r]{$#1^{#4}$}\makebox[5.2em][l]{\,{\scriptsize$[#2,#3]$}}}

\usepackage{graphicx}
\usepackage{epigraph}
\usepackage[table]{xcolor}
\usepackage{pgf}

\newcommand{\gradcell}[1]{%
  \begingroup
  \pgfmathsetmacro{\val}{#1}%
  \def\cellshade{}%
  \ifdim\val pt>0pt
    \pgfmathtruncatemacro{\shade}{min(80,round(80*\val/0.75))}%
    \xdef\cellshade{\noexpand\cellcolor{poscolor!\shade!white}}%
  \else
    \ifdim\val pt<0pt

    \pgfmathtruncatemacro{\shade}{min(80,round(80*abs(\val)/0.75))}%
      \xdef\cellshade{\noexpand\cellcolor{negcolor!\shade!white}}%
    \fi
  \fi
  \endgroup
  \cellshade #1%
}

\definecolor{negcolor}{HTML}{CD3700}
\definecolor{poscolor}{HTML}{018571}

\newcommand{\para}[1]{\vspace{0.2em}\noindent\textbf{\textit{#1}~}}

\title{Answer Bubbles: Information Exposure in AI-Mediated Search}
\author{
 \textbf{Michelle Huang\textsuperscript{$\ddagger$}},~
 \textbf{Agam Goyal\textsuperscript{$\ddagger$}},~
 \textbf{Koustuv Saha},~
 \textbf{Eshwar Chandrasekharan}
\\
 Siebel School of Computing and Data Science\\  University of Illinois Urbana-Champaign, Urbana, IL, USA
\\
\texttt{\{mh106, agamg2, ksaha2, eshwar\}@illinois.edu}
}

\begin{document}
\maketitle
\def\thefootnote{$\ddagger$}\footnotetext{Both authors contributed equally. Author order decided by Nintendo Super Smash Bros.}
\def\thefootnote{\arabic{footnote}}

\begin{abstract}
Generative search systems are increasingly replacing link-based retrieval with AI-generated summaries, yet little is known about how these systems differ in sources, language, and fidelity to cited material. We examine responses to 11,000 real search queries across five systems---vanilla GPT, Search GPT, Perplexity Search with Grok, Google AI Overviews, and traditional Google Search---at three levels: source diversity, linguistic characterization of the generated summary, and source-summary fidelity. 
We find that generative search systems exhibit significant \textit{source-selection} biases in their citations, favoring certain sources over others.
Incorporating search also selectively attenuates epistemic markers, reducing hedging by up to 60\% while preserving confidence language in the AI-generated summaries.
At the same time, AI summaries further compound the citation biases: Wikipedia and longer sources are disproportionately overrepresented, whereas cited social media content and negatively framed sources are substantially underrepresented.
Our findings highlight the potential for \textit{answer bubbles}, in which identical queries yield structurally different information realities across systems, with implications for user trust, source visibility, and the transparency of AI-mediated information access.\footnote{Code for replicating our experiments can be found in \href{https://github.com/scuba-illinois/answer-bubbles-audit}{https://github.com/scuba-illinois/answer-bubbles-audit}}
\end{abstract}

\section{Introduction}

\epigraph{``Google users are less likely to click on a link when they encounter search pages with AI summaries.''}{---Pew Research Center, July 2025~\cite{Chapekis_Lieb_2025}}

Generative search systems are increasingly mediating how users search for and access information, shifting the landscape from ``traditional'' link-based search toward AI-generated overviews that select, synthesize, and cite sources on users’ behalf~\cite{gao2023retrieval,li2024survey}. Rather than returning ranked lists of links for users to evaluate themselves, these AI systems collapse the search process into a single synthesized output that implicitly determines what information the users receive, how it is framed, and which sources are attributed. 

\begin{figure}[t]
    \centering
    \includegraphics[width=\columnwidth]{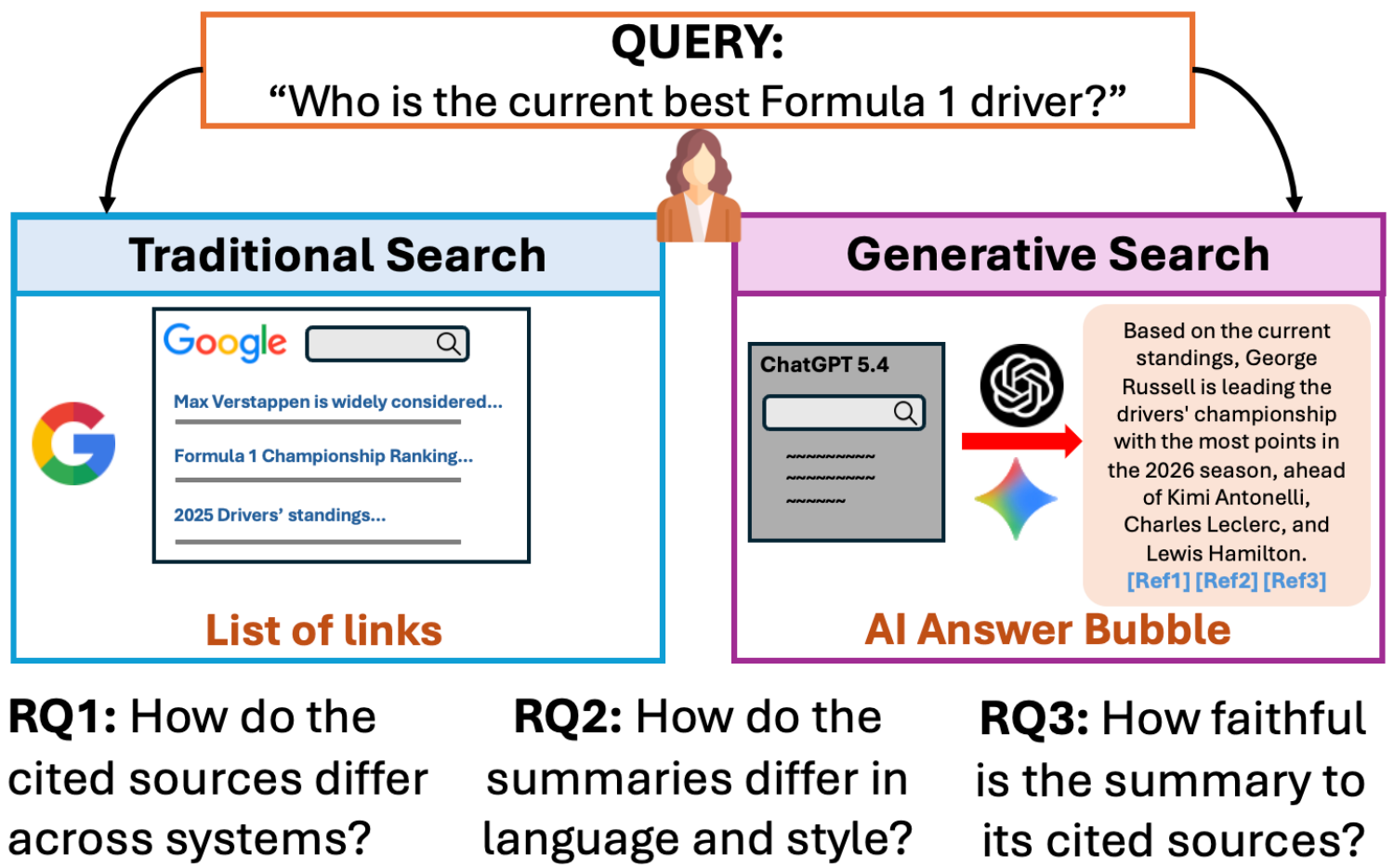}
    \caption{\textbf{Paper Overview.} Traditional search returns a ranked list of links for users to evaluate, while generative search produces an \emph{answer bubble} containing AI-generated summaries synthesized from multiple sources. 
    We study these answer bubbles along three dimensions: the sources they cite (RQ1), the linguistic and epistemic qualities of their summaries (RQ2), and how faithfully those summaries represent cited content (RQ3).}
    \label{fig:teaser}
\end{figure}

Recent evidence suggests that this shift is already reshaping user behavior at scale. A Pew Research study found that when Google's AI Overview appeared in search results, users clicked on traditional result links at roughly half the rate compared to searches without a summary, and clicked on sources cited within the AI summary itself in only 1\% of visits~\cite{Chapekis_Lieb_2025}. As users increasingly rely on these synthesized overviews rather than navigating to original sources, the design choices embedded in generative search systems, i.e., which sources they retrieve, how they synthesize them, and what language they use, can have great impact on how they shape public information exposure~\cite{sharma2024generative}.

Despite this growing influence, systematic empirical evaluations across generative search systems remain limited. Prior work has examined informational accuracy in individual large language model (LLM)-based systems~\cite{vu-etal-2024-freshllms}, but how different systems utilize various types of cited sources, incorporate epistemic attributes in outputs, and maintain fidelity to the original cited information remains understudied. This gap is consequential: just as \emph{filter bubbles} describe how personalized algorithms restrict the diversity of content users encounter~\cite{pariser2011filter}, generative search systems risk creating what we term \textit{``answer bubbles''}: self-contained, system-specific information realities in which the same query issued to different search platforms could yield structurally different answers, without the user's knowledge or ability to compare across systems.

Furthermore, information-seeking is also influenced by \emph{how} it is presented. Affective and epistemic cues in language can signal trustworthiness, curiosity, or authority~\cite{vogl2020surprised, krakowska2020affective}. When synthesizing text, these attributes are no longer properties of the original sources alone, but of the system's own voice. Whether and how these attributes vary across systems remains an open question with direct implications for user trust and information quality.

In this paper, we present the first comparative study of generative and traditional search systems. We query 11{,}000 real user search queries across 11 topics through five systems---vanilla GPT, Search GPT, Perplexity Search with Grok, Google AI Overviews, and traditional Google Search---and ask three research questions (RQs):

\noindent \textbf{RQ1:} How do the sources cited by generative search systems differ from each other and from traditional search in terms of diversity, concentration, and domain composition?

\noindent \textbf{RQ2:} How do the epistemic, psycholinguistic, and stylistic qualities of generative search responses vary across systems and topics?

\noindent \textbf{RQ3:} How selectively do generative search summaries represent information from their sources?

\para{Summary of Findings:} Our analysis yields three key findings. First, generative search systems draw from fundamentally different source pools, with Search GPT's top-100 domains overlapping only 24–25\% with traditional Google Search, favoring encyclopedic and news sources while nearly excluding social media. Second, search grounding systematically reshapes the epistemic character of responses, reducing hedging language by up to 60\% while preserving confidence markers, with effects varying significantly by topic. 
Third, biases in source selection compound during synthesis: Wikipedia is both the most-cited and most over-represented source, whereas social media and discussion forums, despite being cited, are substantively under-used by up to 22 percentage points.
\vspace{-12pt}
\section{Related Work}

\para{Auditing Generative Search Systems:} The shift from link-based retrieval to LLM-synthesized search responses has prompted growing scrutiny of how these systems access and present information. \citet{liu-etal-2023-evaluating} audited citation verifiability across four generative search engines, finding that only 51.5\% of generated statements were fully supported by their citations, and \citet{narayanan2025search} similarly identified 16 recurring limitations, including frequent hallucination and inaccurate citation through a combination of a user study and automated evaluation. \citet{li2024generative} conducted a smaller-scale audit of ChatGPT, Bing Chat, and Perplexity, documenting sentiment and source-authority biases across public-interest topics. Domain-specific audits have surfaced concrete downstream harms: \citet{hu2026auditing} found that Google's AI Overviews and Featured Snippets disagree with each other on 33\% of baby care and pregnancy queries and rarely include medical safeguards. At larger scale, \citet{xu2026measuring} measured AI Overview activation, source quality, and claim fidelity across 55,393 trending queries, finding that activation is highly sensitive to query phrasing and that nearly 30\% of AIO-cited domains never appear among the corresponding organic results, suggesting AIO draws from a source pool distinct from traditional ranking. \citet{vu-etal-2024-freshllms} showed that augmenting LLMs with search improves factual accuracy on time-sensitive queries, and \citet{li2024survey} surveyed LLM-powered search and recommendations. However, these studies focus on individual system properties or a single platform---accuracy, verifiability, bias, or one product's activation behavior---rather than systematically comparing user-relevant properties across systems. \textit{Our work fills this gap with a large-scale cross-system study spanning source selection, linguistic characterization, and source-summary coverage.}

\para{Source Bias in Retrieval and Generation:} Search engines have long shaped source visibility through ranking concentration~\cite{trielli2019search}. Generative search deepens this dynamic, as systems now actively select which sources to incorporate and suppress~\cite{dai2024neural}. Prior work has shown that dense retrievers exhibit systematic biases toward short, surface-level documents~\cite{coelho-etal-2024-dwell,fayyaz-etal-2025-collapse}. This selectivity extends beyond document length: \citet{samuel2026does} found that equipping rerankers with explicit reasoning neither improves nor worsens demographic fairness relative to non-reasoning rerankers, with geographic fairness gaps persisting regardless of architecture, \citet{jang2026you} showed that adaptive retrieval-augmented generation pipelines are highly sensitive to semantically-equivalent rephrasings of the same query, changing retrieval decisions and answer quality even when user intent is unchanged, and \citet{goyal-etal-2026-masking} showing that these biases may not be mitigated even with advanced query enhancement techniques. Our work extends these findings beyond the retrieval stage, showing that source-selection biases compound during synthesis. \textit{We operationalize this using Equal Coverage and Coverage Parity metrics adapted from~\citet{li2025coverage}, quantifying source-summary coverage as a measurable property of generative search.}

\para{Epistemic and Psycholinguistic Dimensions:} Psycholinguistic factors and epistemic emotions shape how users seek and evaluate information~\cite{krakowska2020affective, vogl2020surprised}, while linguistic register influences trust~\cite{zhao2021more, choo2023climate}. Furthermore, behavioral evidence shows users increasingly rely on AI summaries over original sources~\cite{Chapekis_Lieb_2025}, amplifying the influence of system-generated language. Research on human-AI interaction further suggests users tend to over-rely on AI outputs lacking explicit uncertainty signals~\cite{buccinca2021trust}. \textit{We operationalize these constructs through LIWC-based analyses, showing that search grounding selectively suppresses hedging while preserving confidence markers.}
\section{Data \& Response Collection}

We now outline our process for curation of the query dataset and subsequent response collection.

\para{Query Dataset:} We drew queries from Google's Natural Questions (NQ) corpus, a collection of real user search queries issued to Google with an information-seeking intent~\cite{kwiatkowski-etal-2019-natural}. From this corpus, we randomly sampled 50{,}000 queries and classified each into one of 11 topical categories using GPT-4o-mini in a few-shot setting with temperature 0. The 11 topics (listed in \autoref{tab:query_key_terms}) were adapted from prior work~\cite{huynh2025promptdsi} through a combination of zero-shot topic modeling using BERTopic~\cite{grootendorst2022bertopic} and manual inspection of queries categories. To validate the automated classification by GPT-4o-mini, two authors independently labeled a random subset of 120 queries, achieving 90.00\% and 90.83\% agreement with GPT-4o-mini with an inter-annotator $\kappa = 0.9910$ between the two human raters. To construct the final dataset, we randomly sampled 1{,}000 queries per topic, yielding a balanced set of 11{,}000 queries. See \S\ref{app:appendix} for descriptive statistics and details on key terms in each topic.

\para{Response Collection:} We collected responses to each of the 11{,}000 queries from three sources, yielding 44{,}000 query--response pairs:

\noindent\textbf{(1) Vanilla GPT:} We queried GPT-4o-mini via the OpenAI API without web search. These responses reflect the model's parametric knowledge alone.

\noindent\textbf{(2) Search GPT:} We queried GPT-4o-mini with the \textit{``web\_search''} tool call, which retrieves and cites web sources in its generated responses. These responses represent search-grounded LLM output.

\noindent\textbf{(3) Google AI Overview (Google AIO):} We programmatically issued each query to Google Search via \href{https://serpapi.com/}{SerpAPI} and extracted the AI Overview (a generative summary powered by a Gemini model) if produced. SerpAPI has been used extensively by prior work for scraping Google search results~\cite{li2024large,vu-etal-2024-freshllms,sun2025zerosearch}.

The temperature for GPT generation was set to 0, and the location for querying SerpAPI was a fixed location in Urbana, Illinois, United States. See \S\ref{app:prompts} for response-generation prompts.

To assess external validity beyond OpenAI and Google, we additionally collected responses to the same 11{,}000 queries from Perplexity Search (\href{https://docs.perplexity.ai/docs/search/quickstart}{https://docs.perplexity.ai/docs/search/quickstart}) using model \texttt{xai/grok-4-1-fast-non-reasoning} \cite{xGrok} as the underlying LLM, following the same querying protocol. Since this comparison reinforces our main findings rather than altering them, we report it separately in \S\ref{app:perplexity}.
\section{RQ1: Where Do Various Search Systems Get Their Information?}

In this section, we analyze what types of sources generative and traditional search systems synthesize information from to answer user queries. Across query topics, we aggregate and compare counts of domains and specific websites that were cited by GPT and Google AIO.

\para{Citation volume, coverage, and overlap:} The three systems differ in the number of sources cited. Google Organic returns a near-constant 9--10 results per query (the first page of Google search). Search GPT cites a median of 3 sources, with 97\% of queries containing at least one citation. Google AIO, however, has the most variable coverage, with only 57.8\% of queries triggering a generative overview with a mean of 7.0 references. See \autoref{tab:rq1_summary} for more details. We also find that coverage varies by topic: business (84\%), history (81\%), and education (74\%) most frequently produce AIOs, while sports (34\%), technology (32\%), and travel (32\%) rarely do. To quantify overlap, we computed Jaccard similarity on each pair's top-100 most-cited domains. While Google AIO and Organic share 68\% of their top-100 domains, Search GPT's top-100 overlaps only 24\% with AIO and 25\% with Organic Google Search, indicating that it retrieves from a largely distinct set of sources.

\para{Domain-level source patterns:} Wikipedia is the most cited domain across all systems: Organic Google Search (81\% of queries), Google AIO (28\%), and Search GPT (49\%). Search GPT also surfaces other encyclopedic and news-wire sources including Britannica (9\%), Reuters (5\%), and AP News (4\%). In contrast, Google AIO and Organic prominently feature user-generated platforms: Facebook (10\% and 36\%), Quora (8\% and 24\%), and Reddit (7\% and 42\%).  

Grouping domains into categories, we observe the largest contrast between social media and forums. First, Search GPT draws almost none of its citations from social platforms (0.1\%), while Google AIO (8.5\%) and Organic (13.4\%) rely on them substantially more. Conversely, Search GPT draws disproportionately from encyclopedic and reference sources (27.3\% vs.\ $\sim$10\% for both Google systems) and from news/media outlets (7.2\% vs.\ 2.4\% and 1.7\%). Government and institutional sources are roughly comparable across all three ($\sim$7--9\%). These patterns indicate fundamentally different source-selection philosophy: while Search GPT seems to favor authoritative, editorially-curated content, both of Google's systems surface a broader range of user- and community-driven sources.

\begin{figure*}[t]
\centering
\includegraphics[width=\textwidth]{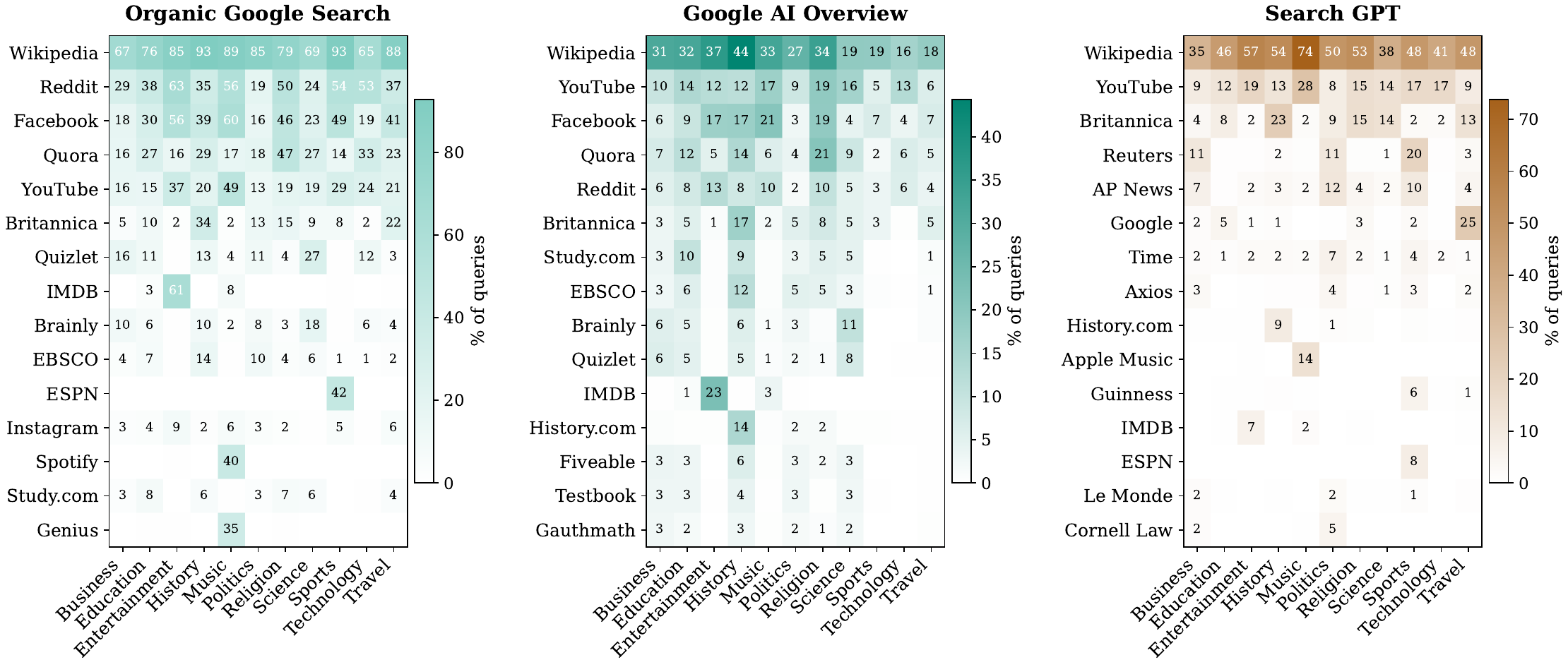}
\caption{Top-15 cited domains by query topic for each source (\% of queries citing each domain). Cell values $\geq$1\% are shown. Domain preferences are strongly topic-dependent: IMDB dominates entertainment, ESPN dominates sports, and Spotify/Genius dominate music, but only in Google's systems and not Search GPT.\vspace{-12pt}}
\label{fig:rq1_domains_topic}
\end{figure*}

\para{Topic-level source distributions:} Domain preferences are strongly topic-dependent (\autoref{fig:rq1_domains_topic}). Certain domains concentrate almost exclusively in specific topics. For example, IMDB dominates entertainment queries (61\% Organic, 23\% AI Overview, 7\% Search GPT), ESPN dominates sports (43\% Organic, 8\% Search GPT), and Spotify (40\%) and Genius (35\%) appear almost exclusively in music queries within Organic results. Search GPT shows distinctive topic-specific information sourcing patterns, with Reuters and AP News concentrated in sports and politics/news (20\% and 12\% respectively), while google.com appears in 25\% of travel queries (likely Google Maps/Travel content). Google AIO surfaces more educational platforms for factual topics---study.com (11\% education), brainly.com (11\% science), and quizlet.com (8\% science)---while Quora reaches 21\% for religion queries. Wikipedia itself varies substantially by topic, ranging from 65\% (technology) to 93\% (history/sports) in Organic results and from 35\% (business) to 74\% (music) in Search GPT.

\section{RQ2: Linguistic Characterization of Generative Responses}

\begin{table*}[t]
\small
\centering
\sffamily
\resizebox{0.9\textwidth}{!}{%
\begin{tabular}{lrrrcccc}
& \multicolumn{3}{c}{\textbf{Mean}} & & \multicolumn{3}{c}{\textbf{Cohen's $d$}} \\
\cmidrule(lr){2-4} \cmidrule(lr){6-8}
\textbf{Feature} & $\textbf{GPT}_\textbf{V}$ & $\textbf{GPT}_\textbf{S}$ & $\textbf{Google}_\textbf{AI}$ & $\textbf{GPT}_\textbf{V}$ vs $\textbf{GPT}_\textbf{S}$ & $\textbf{GPT}_\textbf{V}$ vs $\textbf{Google}_\textbf{AI}$ & $\textbf{GPT}_\textbf{S}$ vs $\textbf{Google}_\textbf{AI}$ \\
\cmidrule(lr){2-4} \cmidrule(lr){6-8}
\rowcolor{gray!15}\multicolumn{8}{l}{\textit{Verbosity \& readability}} \\
Word Count          & 87    & 140   & 83 & \dval{\gradcell{-0.65}}{***} & \dval{\gradcell{0.05}}{***}  & \dval{\gradcell{0.64}}{***}  \\
Avg.\ Sent.\ Length & 16.3  & 18.0  & 11.4 & \dval{\gradcell{-0.17}}{***} & \dval{\gradcell{0.58}}{***}  & \dval{\gradcell{0.59}}{***}  \\
Avg.\ Word Length   & 5.07  & 5.25  & 4.17 & \dval{\gradcell{-0.34}}{***} & \dval{\gradcell{1.06}}{***}  & \dval{\gradcell{1.31}}{***}  \\
Coleman-Liau Index  & 11.5  & 12.8  & 11.7 & \dval{\gradcell{-0.32}}{***} & \dval{\gradcell{-0.06}}{**}  & \dval{\gradcell{0.27}}{***}  \\
Type-Token Ratio    & 0.73  & 0.64  & 0.75 & \dval{\gradcell{0.73}}{***}  & \dval{\gradcell{-0.14}}{n.s.}& \dval{\gradcell{-0.65}}{***} \\
\hdashline
\rowcolor{gray!15}\multicolumn{8}{l}{\textit{Affect}} \\
affect              & 0.054 & 0.053 & 0.031 & \dval{\gradcell{0.03}}{n.s.} & \dval{\gradcell{0.57}}{***}  & \dval{\gradcell{0.58}}{***}  \\
posemo              & 0.045 & 0.044 & 0.025 & \dval{\gradcell{0.04}}{n.s.} & \dval{\gradcell{0.61}}{***}  & \dval{\gradcell{0.61}}{***}  \\
negemo              & 0.008 & 0.009 & 0.006 &  \dval{\gradcell{-0.02}}{***} & \dval{\gradcell{0.14}}{***}  & \dval{\gradcell{0.17}}{***}  \\
\hdashline
\rowcolor{gray!15}\multicolumn{8}{l}{\textit{Cognition}} \\
cogproc             & 0.087 & 0.067 & 0.044 & \dval{\gradcell{0.39}}{***}  & \dval{\gradcell{0.77}}{***}  & \dval{\gradcell{0.48}}{***}  \\
insight             & 0.021 & 0.018 & 0.010 & \dval{\gradcell{0.15}}{***}  & \dval{\gradcell{0.55}}{***}  & \dval{\gradcell{0.44}}{***}  \\
tentat              & 0.025 & 0.015 & 0.010 & \dval{\gradcell{0.43}}{***}  & \dval{\gradcell{0.63}}{***}  & \dval{\gradcell{0.27}}{***}  \\
compare             & 0.028 & 0.023 & 0.016 & \dval{\gradcell{0.19}}{***}  & \dval{\gradcell{0.51}}{***}  & \dval{\gradcell{0.39}}{***}  \\
\hdashline
\rowcolor{gray!15}\multicolumn{8}{l}{\textit{Analytical}} \\
cause               & 0.018 & 0.018 & 0.011 & \dval{\gradcell{0.02}}{***}  & \dval{\gradcell{0.34}}{***}  & \dval{\gradcell{0.35}}{***}  \\
certain             & 0.010 & 0.007 & 0.005 & \dval{\gradcell{0.21}}{***}  & \dval{\gradcell{0.41}}{***}  & \dval{\gradcell{0.24}}{***}  \\
discrep             & 0.006 & 0.003 & 0.002 & \dval{\gradcell{0.26}}{***}  & \dval{\gradcell{0.37}}{***}  & \dval{\gradcell{0.14}}{***}  \\
differ              & 0.021 & 0.013 & 0.011 & \dval{\gradcell{0.36}}{***}  & \dval{\gradcell{0.45}}{***}  & \dval{\gradcell{0.13}}{***}  \\
\hdashline
\rowcolor{gray!15}\multicolumn{8}{l}{\textit{Feature-based model scores}} \\
Politeness          & 0.52  & 0.45  & 0.41  & \dval{\gradcell{0.39}}{***}  & \dval{\gradcell{0.65}}{***}  & \dval{\gradcell{0.30}}{***}  \\
Formality           & 0.93  & 0.95  & 0.84  & \dval{\gradcell{-0.16}}{***} & \dval{\gradcell{0.80}}{***}  & \dval{\gradcell{0.99}}{***}  \\
Polarity            & 0.10  & 0.09  & 0.09  & \dval{\gradcell{0.05}}{n.s.} & \dval{\gradcell{0.09}}{***}  & \dval{\gradcell{0.05}}{***}  \\
Subjectivity        & 0.39  & 0.39  & 0.40  & \dval{\gradcell{0.00}}{*}    & \dval{\gradcell{-0.05}}{n.s.}& \dval{\gradcell{-0.06}}{***} \\
\bottomrule
\end{tabular}}
\caption{Psycholinguistic and lexico-semantic differences across response sources (Benjamini-Hochberg corrected). KW: Kruskal-Wallis test. Pairwise significances determined by Mann-Whitney $U$. \footnotesize ***$p<.001$, **$p<.01$, *$p<.05$.\vspace{-12pt}}
\label{tab:rq2_main}
\end{table*}

Having established \emph{where} generative search systems obtain their information (RQ1), we now examine \emph{how} they present it. To characterize the linguistic qualities of the 33,000 generated responses, we computed a range of lexical, stylistic, and model-based features spanning three categories:

\noindent\textbf{(1) LIWC categories:} We scored each response on 73 Linguistic Inquiry and Word Count (LIWC) categories~\cite{pennebaker2015development}. We focus on three conceptual groupings: \emph{Affect} (positive and negative emotion), \emph{Cognition} (insight, tentative, comparative, and overall cognitive process language), and \emph{Analytical} (causal, certainty, discrepancy, and differentiation language). Scores are expressed as proportions of total word count.

\noindent\textbf{(2) Verbosity and readability:} We measured \emph{word count}, \emph{average sentence length}, \emph{average word length}, and \emph{type-token ratio} (TTR), which is the ratio of unique word types to total tokens, indicating lexical diversity. We assessed readability through the Coleman-Liau index~\cite{Ward2026textstat}.

\noindent\textbf{(3) Model-based scores:} We computed four feature-based scores using transformer-based models. \emph{Politeness} was measured with a fine-tuned XLM-RoBERTa classifier~\cite{srinivasan2022tydip}, outputting the probability of a text being polite (0--1).
\emph{Formality} was measured with a fine-tuned RoBERTa classifier~\cite{babakov2023dont}, outputting a score from 0 (informal) to 1 (formal).
\emph{Polarity} quantifies sentiment from $-1$ (negative) to $+1$ (positive), via TextBlob~\cite{Loria2026sloria}.
\emph{Subjectivity} measures the extent to which text reflects personal opinion (0--1), also via TextBlob. 

These linguistic measures have been used widely by prior research on auditing outputs of generative models~\cite{10.1145/3709366,goel2026inform} and analysis of content arising from online interactions~\cite{pal2026hidden,goyal2026language}. To compare the three response sources (GPT-Vanilla, GPT-Search, and Google AIO), we used Kruskal-Wallis H tests for each feature, followed by pairwise Mann-Whitney U tests with Benjamini-Hochberg correction~\cite{benjamini1995controlling} to determine statistical significance. Effect sizes are reported as Cohen's $d$.

In comparing linguistic characteristics of the generative responses, the key design variable is search grounding: Vanilla GPT generates from parametric knowledge alone, whereas Search GPT and Google AIO both retrieve web content before generation. ~\autoref{tab:rq2_main} reports all features across these systems. Three major patterns emerge: 

\para{Structural divergence:} Despite all being LLM-based, the three systems produce characteristically different responses. The largest contrast is between the two search-grounded systems: as compared to Google AIO, Search GPT responses are 69\% longer ($M\!=\!140$ vs.\ 83 words; $d\!=\!0.64$), use longer words ($d\!=\!1.31$), are substantially more formal ($d\!=\!0.99$), and infer a higher reading level (Coleman-Liau: $M\!=\!12.8$ vs.\ 11.7 grade level; $d\!=\!0.27$). In contrast, Google AIO favors shorter sentences ($M\!=\!11.4$ vs.\ 18.0 words; $d\!=\!0.59$), simpler vocabulary, and higher lexical diversity ($d\!=\!0.65$). These differences suggest completely different design philosophies. While Search GPT produces elaborate, paragraph-style information, Google AIO produces compressed, content-heavy information optimized for quick consumption within a search interface. Vanilla GPT falls between the two on verbosity ($M\!=\!87$ words), shares Search GPT's syntactic complexity and formality, and has TTR comparable to Google AIO ($d\!=\!-0.14$) but much higher than Search GPT ($d\!=\!0.73$). Therefore, Vanilla GPT achieves conciseness through vocabulary variety rather than structural compression.

\para{Epistemic and emotional attenuation:} The systems exhibit significant epistemic differences.

\textbf{(1) Hedging and uncertainty:} Tentative language (\emph{``maybe,'' ``perhaps''}), the primary marker of epistemic caution, drops by 40\% from Vanilla GPT to Search GPT ($0.025 \to 0.015$; $d\!=\!0.43$) and by 60\% from Vanilla GPT to Google AIO ($0.025 \to 0.010$; $d\!=\!0.63$). Overall cognitive process language also drops by 23\% and 49\%, respectively, indicating that hedging is reduced at a faster rate than other types of cognitive language.

\textbf{(2) Reasoning depth:} Insight language (\emph{``think,'' ``know''}) is halved from Vanilla GPT to Google AIO ($0.021 \to 0.010$; $d\!=\!0.55$), and comparative language (\emph{``versus,'' ``rather''}) follows a similar pattern ($d\!=\!0.51$). Analytically, causal language (\emph{``because,'' ``therefore''}) and certainty markers (\emph{``always,'' ``never''}) also decline, though with smaller effects ($d\!=\!0.34$ and $d\!=\!0.41$, respectively). Notably, certainty drops slower than hedging, where the ratio of certainty-to-tentative language (which measures assertiveness) rises from 0.39 in Vanilla GPT to 0.49 in both search systems. Therefore, search grounding does not just compress all epistemic language equally, but rather selectively removes the markers of deliberation while preserving markers of confidence.

\textbf{(3) Emotional flattening:} Positive emotion drops by 44\% from Vanilla GPT to Google AIO ($0.045 \to 0.025$; $d\!=\!0.61$), while negative emotion drops by only 25\% ($d\!=\!0.14$). This asymmetry means Google AIO's affect is not just reduced, but shifted to a lower positive-to-negative emotion ratio. The ML-based polarity and subjectivity scores, by contrast, show negligible differences ($d \leq 0.09$), indicating that the emotional flattening operates at the lexical level, rather than the semantic level, by using fewer emotion words. Google AIO's politeness is also markedly lower than Vanilla GPT's ($d\!=\!0.65$), suggesting a more impersonal style.

\para{Topic-wise variation:} These effects are not uniform across topics. Technology and travel/geograp\-hy show the largest divergence (mean $|d|\!>\!1.0$), with Google AIO reducing 64\% of hedging and 74\% of positive emotion relative to Search GPT ($d\!=\!1.37$ for cogproc, $d\!=\!1.30$ for posemo) in technology queries. Notably, these two topics also least frequently generated AIOs (32\% availability per RQ1). At the other extreme, education, history, and business/finance show near-zero epistemic divergence ($d\!=\!-0.03$ to $0.12$ for hedging and cogproc), indicating that the two search-grounded systems produce effectively indistinguishable output for these high AI-overview-availability topics (74--84\%). Interestingly, for politeness, politics/news and religion show nearly no differences across search systems ($d\!=\!0.02$ and $d\!=\!0.00$), meaning Google AIO matches Search GPT's courtesy for more socially sensitive queries. This contrasts with technology ($d\!=\!0.84$) and sports ($d\!=\!0.94$), where AIOs tend to adopt a much terser tone. Formality, in contrast, remains high across nearly all topics ($d\!=\!0.82$--$1.76$), suggesting very low topical sensitivity. See \autoref{tab:rq2_topic_heatmap} in \S\ref{app:appendix} for detailed results.

\section{RQ3: Source-Summary Coverage}

We now shift focus from how generative search systems \emph{cite} sources (RQ1) and \emph{express} information (RQ2) to how selectively they \emph{synthesize} cited content. Specifically, we examine whether generative summaries draw evenly from sources or systematically favor certain sources over others. Our goal is \textit{descriptive}: these metrics characterize which cited sources are reflected in a summary, but they do not determine whether an under-covered source is lower quality, redundant, or less relevant.

\noindent\textbf{Methods.} We adapt the coverage-based fairness framework of \citet{li2025coverage} to the generative search setting. The pipeline consists of three steps:

\noindent\textbf{(1) Atomic Content Unit (ACU) decomposition:} Each generative response is decomposed into ACUs (single, self-contained factual statements) using GPT-4o-mini with few-shot prompting. We generated ACUs for 1{,}100 queries per source (100 per topic), producing 17{,}644 ACUs for Search GPT (mean 16.0 per summary) and 16{,}898 for Google AIO (mean 15.4). See \S\ref{app:prompts} for ACU prompts.

\noindent\textbf{(2) Extract source context:} For each query, we scraped the full textual content of every cited URL using \texttt{trafilatura}~\cite{barbaresi-2021-trafilatura} and \texttt{BeautifulSoup4}, filtering for core article content excluding ads and boilerplate. Each source document is chunked into overlapping blocks of approximately 100 words, following \citet{li2025coverage}. 

\noindent\textbf{(3) Entailment probability:} We compute entailment probabilities $P(\text{chunk entails ACU})$ for all (source chunk, ACU) pairs using RoBERTa-large fine-tuned on MNLI~\cite{liu2019roberta}. ACUs with mutual probability $>0.95$ are deduplicated via connected-component clustering~\cite{li2025coverage}. For each source and deduplicated ACU group, we take the maximum entailment probability over that source's chunks, then average across covered ACU groups to obtain the source-level coverage probability $p(q,k)$. This max-over-chunks aggregation asks whether any part of a source supports a summary claim; we discuss its length sensitivity in the Limitations. We compute two coverage metrics:

\textbf{(i) Equal Coverage (EC):} For a single query, EC measures whether the summary draws evenly from all $K$ cited sources:
\begin{equation}
\text{EC}(q) = \frac{1}{K}\sum_{k=1}^{K} \left| p(q, k) - \bar{p}(q) \right|
\end{equation}
where $p(q, k)$ is the source-level coverage probability for source $k$ after max-over-chunks aggregation and averaging across covered ACU groups, and $\bar{p}(q) = \frac{1}{K}\sum_k p(q,k)$ is the overall mean across sources. $\text{EC}=0$ indicates perfectly even coverage, with higher values indicating selective reliance. 

\textbf{(ii) Coverage Parity (CP):} CP tests whether sources with a particular attribute $a$ are systematically over- or under-represented across the corpus:
\begin{equation}
\text{CP} = \frac{1}{|A|}\sum_{a \in A} \left| \mathbb{E}_q\left[ p(q,k) - \bar{p}(q) \mid \text{attr}(k) = a \right] \right|
\end{equation}
Lower CP values indicate less systematic bias. We adapt CP to the generative search setting by using attribute schemes that capture structural, topical, and linguistic properties of cited sources. We assess statistical significance via bootstrap resampling (1{,}000 iterations) and use the Mann-Whitney $U$ test for cross-source comparison.

\noindent\textbf{Findings.} \autoref{tab:rq3_ec} reports per-topic EC and coverage probability, and \autoref{tab:rq3_cp} reports CP across attribute schemes. 

\para{Google AIO covers sources less evenly than Search GPT:} Google AIO exhibits significantly higher EC than Search GPT ($M=0.177\pm0.091$ vs. $0.164\pm0.113$; $p<.001$), indicating that it distributes attention across cited sources less evenly. Google AIO also achieves lower overall coverage (0.447 vs.\ 0.500; $p<.001$), meaning its summaries capture less of the total content in cited sources despite citing nearly twice as many (mean$=$5.0 vs. 2.7). This suggests a trade-off: Google AIO cites more sources but synthesizes each one less thoroughly. Topic-wise, for Search GPT, music shows the highest coverage inequality ($\text{EC}$$=$0.236) and technology the lowest (0.131). For Google AIO, entertainment shows the highest inequality (0.217) while science shows the lowest (0.149). 

Since Google AIO triggers for a different subset of queries than Search GPT, we also restrict the comparison to the 338 queries where both systems produced a valid, scoreable EC. On this matched subset, the EC gap narrows ($0.159$ vs.\ $0.166$) and is no longer significant (Wilcoxon signed-rank $p=0.31$), indicating that part of the aggregate EC gap reflects which queries elicit an AI Overview rather than a stable per-query fidelity difference. The coverage gap, however, remains large and significant on the same matched subset ($0.500$ vs.\ $0.440$; $p<.001$), so Search GPT's higher extraction of cited content is not explained by query selection alone.

\para{Longer sources are the most over-represented:} The largest CP effect for both systems is for source document length (Search GPT$=$0.058; Google AIO$=$0.079). Short sources are strongly under-represented (Search GPT: $-13.4$pp; Google AIO: $-17.7$pp; both $p<.001$), while long sources are over-represented (3.9pp and 4.6pp). This suggests that these systems place more emphasis on the amount of extractable text than on relevance or quality, exhibiting a length bias that parallels brevity biases in dense retrievers~\cite{fayyaz-etal-2025-collapse}.

\para{Wikipedia/encyclopedia/government sources:} Encyclopedic sources are over-represented in both Search GPT ($+2.2$pp) and Google AIO ($+5.0$pp). For Wikipedia specifically (Search GPT: $+2.6$pp; Google AIO: $+5.4$pp; both $p<.001$), we find a significant compounding effect where it is not only cited most often (RQ1), but its content also receives disproportionate weight in summaries. In contrast, government sources are under-represented in both systems ($-4.2$pp and $-3.5$pp, respectively).

\begin{table}[t]
\small
\centering
\sffamily
\resizebox{0.9\columnwidth}{!}{%
\begin{tabular}{lcccc}
& \multicolumn{2}{c}{\textbf{Search GPT}} & \multicolumn{2}{c}{\textbf{Google AIO}} \\
\cmidrule(lr){2-3} \cmidrule(lr){4-5}
\rowcolor{blue!10}\textbf{Topic} & EC & Coverage & EC & Coverage \\
\midrule
\rowcolor{gray!10}Business/Fin./Econ.    & 0.162 & 0.500 & 0.162 & 0.431 \\
Education/Lit.         & 0.189 & 0.481 & 0.175 & 0.465 \\
\rowcolor{gray!10}Entertainment          & 0.185 & 0.480 & 0.217 & 0.459 \\
History                & 0.160 & 0.460 & 0.169 & 0.445 \\
\rowcolor{gray!10}Music & 0.236 & 0.449 & 0.177 & 0.407 \\
Politics/News          & 0.152 & 0.506 & 0.168 & 0.444 \\
\rowcolor{gray!10}Religion               & 0.138 & 0.523 & 0.165 & 0.465 \\
Science                & 0.141 & 0.543 & \textbf{\textcolor{cyan}{0.149}} & 0.453 \\
\rowcolor{gray!10}Sports                 & 0.164 & 0.542 & 0.204 & 0.455 \\
Technology & \textbf{\textcolor{cyan}{0.131}} & 0.535 & 0.189 & 0.433 \\
\rowcolor{gray!10}Travel/Geography       & 0.162 & 0.477 & 0.177 & 0.421 \\
\midrule
\textit{Overall}       & \textit{0.164} & \textit{0.500} & \textit{0.177} & \textit{0.447} \\
\bottomrule
\end{tabular}}
\caption{Equal Coverage (EC) and mean coverage probability (Coverage) by topic. EC measures how unevenly a summary draws from its cited sources (\textbf{\textcolor{cyan}{lower}} = more even). Coverage indicates the fraction of source content entailed by summary ACUs.\vspace{-12pt}}
\label{tab:rq3_ec}
\end{table}

\begin{table}[t]
\small
\centering
\sffamily
\resizebox{0.9\columnwidth}{!}{%
\begin{tabular}{@{}l@{}cc}
\rowcolor{blue!10}& \textbf{Search GPT} & \textbf{Google AIO} \\
\rowcolor{blue!10}& \footnotesize Mean Diff (pp) & \footnotesize Mean Diff (pp) \\
\midrule
\rowcolor{gray!15}\multicolumn{3}{@{}l}{\textit{Document Length}} \\
\quad Short ($<$200 words)   & \dval{-0.134}{***} & \dval{-0.177}{***} \\
\quad Medium (200--800)      & \dval{+0.000}{}    & \dval{+0.015}{**}  \\
\quad Long ($>$800 words)    & \dval{+0.039}{***} & \dval{+0.046}{***} \\[2pt]
\rowcolor{gray!15}\multicolumn{3}{@{}l}{\textit{Source Sentiment}} \\
\quad Positive               & \dval{+0.004}{}    & \dval{+0.011}{**}  \\
\quad Neutral                & \dval{-0.008}{}    & \dval{-0.010}{}    \\
\quad Negative               & \dval{-0.037}{}    & \dval{-0.138}{***} \\[2pt]
\rowcolor{gray!15}\multicolumn{3}{@{}l}{\textit{Domain Category}} \\
\quad Encyclopedia           & \dval{+0.022}{***} & \dval{+0.050}{***} \\
\quad Government             & \dval{-0.042}{*}   & \dval{-0.035}{**}  \\
\quad Social/Forum           & {--}               & \dval{-0.221}{***} \\
\quad Other                  & \dval{-0.015}{*}   & \dval{+0.001}{}    \\[2pt]
\rowcolor{gray!15}\multicolumn{3}{@{}l}{\textit{Wiki vs.\ Non-Wiki}} \\
\quad Wikipedia              & \dval{+0.026}{***} & \dval{+0.054}{***} \\
\quad Non-Wikipedia          & \dval{-0.017}{**}  & \dval{-0.009}{**}  \\[2pt]
\rowcolor{gray!15}\multicolumn{3}{@{}l}{\textit{Citation Position}} \\
\quad Position 0 (first)     & \dval{+0.015}{*}   & \dval{+0.032}{***} \\
\quad Position 1             & \dval{+0.001}{}    & \dval{+0.020}{*}   \\
\quad Position 2             & \dval{-0.021}{*}   & \dval{+0.012}{}    \\[2pt]
\rowcolor{gray!15}\multicolumn{3}{@{}l}{\textit{Source Subjectivity}} \\
\quad High subjectivity      & \dval{-0.023}{*}   & \dval{-0.032}{***} \\
\quad Low subjectivity       & \dval{+0.004}{}    & \dval{+0.008}{**}  \\[2pt]
\bottomrule
\multicolumn{3}{@{}l}{\footnotesize ***\,$p<.001$, **\,$p<.01$, *\,$p<.05$.}\\
\multicolumn{3}{@{}l}{\footnotesize Positive = over-represented; negative = under-represented.}\\
\end{tabular}}
\caption{Coverage Parity (CP) analysis across attribute schemes. Higher CP values indicate more systematic bias. Significance via bootstrap resampling (n=1{,}000). 95\% bootstrap confidence intervals for these estimates are reported in \autoref{tab:rq3_cp_appendix}.\vspace{-12pt}}
\label{tab:rq3_cp}
\end{table}

\para{Negative emotion and social media forums are under-represented:} Source sentiment shows markedly different patterns across systems. Sentiment effects are non-significant for Search GPT (CP $=0.016$) but large and significant for Google AIO (CP $=0.053$): negative-sentiment sources are under-covered by 13.8pp ($p<.001$), while positive-sentiment sources are slightly over-represented ($+1.1$pp, $p<.001$). This asymmetry suggests that Google AIO's synthesis implicitly filters against negative content.

Domain-category analysis further shows that social media and forum sources (Reddit, Quora) are under-covered by 22.1pp in Google AIO ($p < .001$), suggesting that when Google AIO draws far less content from Reddit when cited alongside other sources. Combined with our finding from RQ1 that Google AIO \emph{does} cite social platforms (8.5\% of domains), this highlights a gap between citation and synthesis: the model lists user-driven sources in its references but draws disproportionately little content from them. Reddit ($n=149$, diff~$= -0.221$, $p < .001$) and Khan Academy ($n=21$, diff~$= -0.311$, $p < .001$) are the most under-covered in the synthesized summaries.

\para{Sources with epistemic markers receive more coverage:} Source-level LIWC analysis reveals that sources containing causal language (``because,'' ``therefore'') and certainty markers (``always,'' ``definitely'') are systematically over-covered in both systems. Causal language shows $\text{CP}$=0.071 (Search GPT) and 0.070 (Google AIO). Conversely, highly subjective sources are under-covered in both systems (-2.3pp and -3.2pp). Together, these patterns suggest that the models favor sources with assertive, explanatory prose that is easier to extract factual claims from. 
\section{Discussion}

Our study reveals that generative search systems do not simply retrieve and present information, but they actively reconstruct it through source selection, linguistic transformation, and selective synthesis. We organize our discussion around three interconnected themes that emerge from  our findings.

\para{The Compounding Effect From Citation to Synthesis:} A central finding of this work is that biases in source selection (RQ1) are amplified rather than mitigated during synthesis (RQ3). For example, Wikipedia is not only the most frequently cited domain across all systems, but its content is disproportionately over-represented in generated summaries. Conversely, social media and forum sources present an inverse pattern: Google AIO cites Reddit and Quora in 7–8\% of queries, yet draws 22.1\% less content from social platforms than from other sources. This creates a \emph{citation-synthesis gap}: the model references user-generated sources lending an appearance of source diversity, while  relying on a narrower set of encyclopedic or authoritative texts. This compounding dynamic resonates with concerns raised by \citet{dai2024bias,dai2024neural} about systematic biases in how models prioritize sources, and extends them by showing that the bias operates not only at the retrieval stage but deepens during generation. With the rise of agentic discourse on the internet~\cite{goyal2026social,feng2026moltnet}, compounded effects of biased retrieval towards agent-generated content~\cite{dai2024neural} with the citation-synthesis gap we found can lead to drastic outcomes. For users, this gap is largely invisible, and the cited sources in an AI Overview may suggest breadth that the synthesized text does not reflect. This also has implications for source visibility and the economic viability of smaller publishers and community-driven platforms.

\para{Epistemic Reshaping of Summaries from Uncertain Sources:} Our linguistic analysis (RQ2) reveals that search grounding does not uniformly compress language but selectively reshapes its epistemic character. The ratio of certainty-to-tentative language rises from 0.39 in Vanilla GPT to 0.49 in both search-grounded systems, connecting directly to observations from RQ3 findings that sources containing causal and certainty language are systematically over-covered, while highly subjective sources are under-covered. Together, these results suggest that generative search systems preferentially extract from assertive, explanatory prose and then present the resulting synthesis in a style that is more assertive and less hedged than the underlying sources warrant. This epistemic flattening is particularly concerning as it could potentially present uncertain information to users with unwarranted confidence. This aligns with broader concerns in information-seeking literature about how linguistic style influences perceived credibility~\cite{zhao2021more,choo2023climate} and how epistemic emotions shape knowledge exploration~\cite{vogl2020surprised}. If users interpret the absence of hedging as a signal of factual reliability, the epistemic reshaping we find could miscalibrate the trust-reliance association in educational settings~\cite{10.1145/3772363.3799386}.

\para{Implications and Future Directions:} Our findings surface important implications for different stakeholders. \emph{For NLP researchers}, we offer a reusable framework for auditing source-summary fidelity in any retrieval-augmented generation system. Our finding that document length is the strongest predictor of over-representation  parallels recent work on brevity biases in dense retrievers~\cite{fayyaz-etal-2025-collapse} and suggests that length bias may persist through the full RAG pipeline. Future work should investigate whether debiasing at the retrieval stage propagates to more equitable synthesis, or whether generation introduces its own independent biases. 
Another open question is whether these dynamics generalize to platform-native generative search systems, where retrieval is based on community discourse rather than the open web, making source selection consequential for which contributions within a community are amplified.
\emph{For system designers}, the  divergence between Search GPT and Google AIO demonstrates that these systems have fundamentally different design choices about what constitutes a good search response. Making these design philosophies more transparent to users through source diversity indicators or  confidence signals could help users make more informed judgments about the information they receive. \emph{For users and policymakers}, the key takeaway is that the same query yields different ``answer bubbles'' that the end user is unaware of. We propose that  generative search systems should be subject to transparency requirements analogous to those  for algorithmic recommendation systems~\cite{Kowald2024TransparencyPAA}.
\section{Conclusion}

We presented a large-scale comparative study of generative and traditional search systems across 11{,}000 real user queries, showing that the same query can produce structurally different source sets, summaries, and linguistic framings depending on the system used. Generative search systems diverge sharply in which sources they cite, how they linguistically frame responses, and how selectively they synthesize cited material, with biases in source selection compounding during generation. As these systems increasingly mediate public information access, our findings underscore the need for greater transparency in how AI-generated search responses are constructed and the sources they prioritize.

\section*{Limitations}

Our study has limitations that outline interesting directions for future work.

\para{Temporal and geographic scope:} Our audit captures behavior of generative search systems at a single point in time from a single location. Generative search systems are updated frequently, and source selection strategies, linguistic patterns, and synthesis behaviors may shift with model updates or policy changes. Similarly, Google search results and AI Overview availability vary by geography, language, and user profile. Longitudinal and cross-regional studies are needed to determine how stable the answer bubble patterns we find are.

\para{Query representativeness:} Our queries are drawn from Google's Natural Questions (NQ) corpus, which is primarily focused on information-seeking intent and was originally collected around 2018. Other types of queries may elicit different source selection and linguistic behaviors. Furthermore, our 11-topic list might collapse heterogeneous subtopics into single categories (e.g., ``Science'' spans biology, physics, and healthcare). Finer-grained topic analysis and more diverse query types could reveal additional insights that our current design did not capture.

\para{Methodological proxies:} Our RQ3 pipeline relies on GPT-4o-mini for atomic content unit decomposition and RoBERTa-large for entailment scoring, both of which may introduce noise and domain-specific errors. Since we utilized the work of \citet{li2025coverage}, we did not conduct additional human validation of ACU quality or entailment accuracy. Additionally, not all cited URLs were successfully scraped (e.g., paywalled content), potentially biasing coverage estimates if non-scrapable sources differ systematically from scrapable ones. The max-over-chunks aggregation used to compute source-level coverage also carries a mechanical component, where on a stratified subsample (up to 30 queries/topic), replacing max- with mean-over-chunks substantially shrinks the length-diff pattern in \autoref{tab:rq3_cp} for all three systems (short-source under-representation moves from $-0.148$ to $+0.018$ for Search GPT, $-0.217$ to $-0.011$ for Google AIO, and $-0.145$ to $+0.007$ for Perplexity), so part of the reported length bias reflects the pooling operator rather than length alone. Aggregate EC and CP are comparatively stable across chunk sizes (60/80/120 words), so this sensitivity is specific to the pooling choice rather than the chunking scheme. EC and CP also measure selective coverage rather than normative source quality: they do not adjudicate whether a source deserved more or less representation, whether uncovered material was redundant, or whether a source was higher or lower quality.

\para{No user-study, ideology, or causal claims:} While we provide detailed characterization of  outputs, we do not measure user perception or behavior. Therefore, we cannot make claims about whether users would notice or be affected by the epistemic and source-diversity differences we find. We also do not directly measure ideological skew, political worldview effects, or personalization; our ``answer bubble'' framing refers to system-level differences in source exposure and synthesis. Controlled user studies measuring how answer bubbles influence trust calibration and downstream decision-making, as well as targeted audits of politically charged queries with stance-annotated sources, would directly extend this work. Moreover, our study is correlational, drawing associations between search grounding and linguistic attenuation. We therefore cannot establish causal mechanisms or disentangle the contributions of model architecture, training data, and retrieval pipeline on the effects we find.

\para{System coverage and confounds:} We audit five systems, but the generative search landscape includes many other systems and open-source search-augmented models that may show different results. Additionally, we treat Google AIO as a black box without access to its specific model version, retrieval pipeline, or post-processing steps, limiting mechanistic interpretation of our findings. Future work incorporating a broader set of systems would strengthen the generalizability of our findings.

\section*{Acknowledgments}

This work used the Delta system at the National Center for Supercomputing Applications through allocation \#240481 from the Advanced Cyberinfrastructure Coordination Ecosystem: Services \& Support (ACCESS) program, which is supported by National Science Foundation grants \#2138259, \#2138286, \#2138307, \#2137603, and \#2138296.

\section*{AI Involvement Disclosure}

The research conducted in our work used artificial intelligence tools (e.g., Grammarly, ChatGPT) solely for improving the grammar and readability of certain sections of the manuscript, and for partial development of the codebase for the evaluations conducted through Claude Code, Codex, and Cursor. All scientific content, interpretations, and decisions reflect the original work, judgment and intellectual contributions of the research team. 
\section*{Ethical Considerations}

Our study analyzes system-generated outputs from publicly accessible search  platforms and does not involve human subjects, personal data, or user  interaction. All queries are drawn from the publicly available Natural Questions (NQ) corpus. We followed the terms of service and standard usage policies for both SerpAPI and OpenAI API.

We also want to highlight that we do not intend to single out any particular system with our findings, but aim to characterize broad patterns across the generative search landscape that have implications for information access. We believe that transparent, independent auditing of  AI-mediated search systems serves the public interest.

\bibliography{references}

\appendix
\appendix
\section{Appendix}\label{app:appendix}
\setcounter{table}{0}
\setcounter{figure}{0}
\renewcommand{\thetable}{A\arabic{table}}
\renewcommand{\thefigure}{A\arabic{figure}}

In this section we extend the description of the query dataset by providing simple descriptive statistics and highlighting key terms across topics.

The queries have a mean length of $9.32 \pm 1.91$ words. The majority of queries are phrased as questions (78\%), with the remainder being declarative statements or keyword fragments. The five most frequent key terms across all queries are \textit{world} (377), \textit{India} (346), \textit{name} (338), \textit{time} (327), and \textit{United States} (249). \autoref{tab:query_key_terms} presents the top key terms per topic, computed after lemmatization and aggregating counts across morphological variants.

\begin{table*}[ht]
\small
\sffamily
\centering
\resizebox{\textwidth}{!}{
\begin{tabular}{lrl}
\rowcolor{blue!10}\textbf{Topic} & \textbf{\# Queries} & \textbf{Top 5 Key Terms} \\
\midrule
\rowcolor{gray!15}Business, Finance, \& Economy & 1,000 & company (53), India (51), United States (48), world (43), bank (43) \\
Education \& Literature       & 1,000 & book (107), wrote (95), school (77), name (50), summary (40) \\
\rowcolor{gray!15}Entertainment                 & 1,000 & play (193), movie (116), season (110), episode (90), film (56) \\
Geography \& Travel           & 1,000 & locate (148), map (90), world (69), state (61), river (60) \\
\rowcolor{gray!15}History                       & 1,000 & war (92), name (58), India (58), world (38), United States (34) \\
Music                         & 1,000 & song (239), sing (216), write (88), love (72), lyrics (47) \\
\rowcolor{gray!15}Politics \& News              & 1,000 & state (147), India (90), court (68), president (65), united (61) \\
Religion                      & 1,000 & Bible (235), book (90), church (81), God (70), catholic (45) \\
\rowcolor{gray!15}Science                       & 1,000 & earth (33), locate (32), cell (24), function (23), name (23) \\
Sports                        & 1,000 & time (142), world cup (95), game (75), team (73), NBA (54) \\
\rowcolor{gray!15}Technology                    & 1,000 & use (111), difference (55), computer (53), Xbox (42), Windows (40) \\
\midrule
\textbf{All Queries}          & \textbf{11,000} & \textbf{world (377), India (346), name (338), time (327), United States (249)} \\
\bottomrule
\end{tabular}}
\caption{\textbf{Query Dataset Summary.} Table representing all 11 topics and the number of queries per topic in our final dataset. We also report the top 5 most frequently appearing key terms per topic with frequencies in parentheses.}
\label{tab:query_key_terms}
\end{table*}

\autoref{tab:rq1_summary} reports extended description of our findings from RQ1, including Perplexity (see \S\ref{app:perplexity} for the full Perplexity + Grok experiments).

\begin{table}[ht]
\small
\centering
\sffamily
\resizebox{\columnwidth}{!}{%
\begin{tabular}{lcccc}
\rowcolor{blue!10}& \textbf{Search GPT} & \textbf{Google AIO} & \textbf{Perplexity} & \textbf{Organic} \\
\midrule
Queries w/ $\geq$1 citation & 97.0\% & 57.8\% & 74.5\% & 99.8\% \\
Mean citations/query        & 3.2    & 4.0    & 5.7    & 9.3   \\
Median citations/query      & 3      & 4      & 6      & 10    \\
Unique domains              & 7,606  & 14,279 & 22,562 & 26,891 \\
\hdashline
\rowcolor{gray!15}\multicolumn{5}{l}{\textit{Domain category (\% of cited domains)}} \\
\quad Encyclopedia/Reference   & 27.3\% & 10.3\% & 13.1\% & 11.0\% \\
\quad News/Media               & 7.2\%  & 2.4\%  & 0.4\%  & 1.7\%  \\
\quad Gov./Institutional       & 7.0\%  & 8.5\%  & 9.1\%  & 7.9\%  \\
\quad Social Media/Forums      & 0.1\%  & 8.5\%  & 0.0\%  & 13.4\% \\
\quad Video/Streaming          & 6.9\%  & 3.7\%  & 8.7\%  & 3.3\%  \\
\quad Organization (.org)      & 8.4\%  & 11.9\% & 12.6\% & 10.9\% \\
\quad Other                    & 43.0\% & 54.0\% & 55.8\% & 51.0\% \\
\bottomrule
\end{tabular}}
\caption{\textbf{Source summary.} \textit{Domain categories} show each category's share of all cited domains.\vspace{-8pt}}
\label{tab:rq1_summary}
\end{table}

\subsection{Perplexity + Grok: A Cross-System Robustness Check}\label{app:perplexity}

To assess external validity beyond OpenAI and Google, we replicate our full RQ1--RQ3 pipeline on a third-party generative search system: Perplexity Search paired with \texttt{xai/grok-4-1-fast-non-reasoning} as the underlying LLM, queried on the same 11{,}000-query benchmark. We summarize results below. Overall, Perplexity reinforces the same directional findings as our main analysis, corroborating the answer-bubbles framing across a third model family and platform.

\begin{figure}[ht]
\centering
\includegraphics[width=\columnwidth]{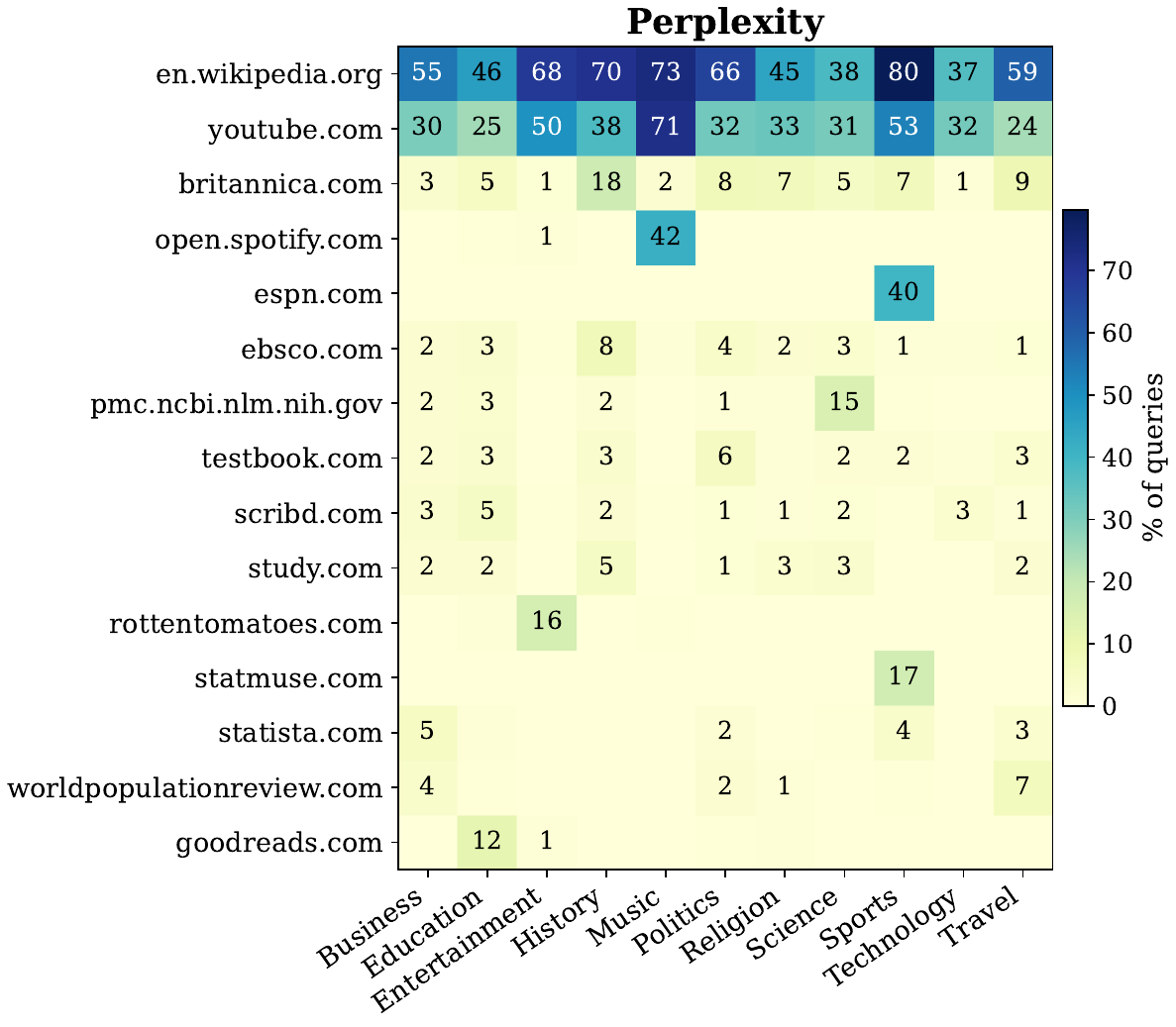}
\caption{Top-15 cited domains by query topic for Perplexity (\% of queries citing each domain), in the same format as \autoref{fig:rq1_domains_topic}. Cell values $\geq$1\% are shown.\vspace{-8pt}}
\label{fig:rq1_perplexity_heatmap}
\end{figure}

\para{RQ1: Source selection.} Perplexity shows 74.5\% query coverage with at least one citation (mean 5.7 citations/query, median 6), placing it between Search GPT (97.0\%, mean 3.2) and Google AIO (57.8\%, mean 4.0), and below Organic Google's near-fixed first-page behavior (99.8\%, mean 9.3); see \autoref{tab:rq1_summary} for the full comparison. Its most-cited domains are dominated by reference and video sources (Wikipedia: 58.0\% of queries; YouTube: 38.1\%), and its domain-category profile remains low in social/forum sourcing (0.0\%) compared with Google AIO (8.5\%) and Organic Google (13.4\%). Top-100 domain overlap is moderate with Google's systems (Jaccard: 0.351 with AIO, 0.379 with Organic), indicating partial convergence but not collapse into Google's retrieval profile. \autoref{fig:rq1_perplexity_heatmap} shows Perplexity's top cited domains by topic, in the same format as \autoref{fig:rq1_domains_topic}: it is especially Wikipedia- and YouTube-heavy in high-fact-density topics such as music (73\% Wikipedia, 71\% YouTube) and sports (80\% Wikipedia, 53\% YouTube), while topic-specific domains still surface (Spotify in music, ESPN in sports, Rotten Tomatoes in entertainment).

\para{RQ2: Linguistic characterization.} We ran Perplexity outputs through the same LIWC and readability/verbosity feature pipeline as \autoref{tab:rq2_main} and compare descriptively in \autoref{tab:rq2_perplexity}. Perplexity is the most verbose system we study (mean 190 words vs.\ 140 for Search GPT and 87 for Vanilla GPT), with readability close to Search GPT (Coleman-Liau 12.85 vs.\ 12.84) but shorter sentences (14.2 vs.\ 18.0 words), suggesting a style that combines long responses with shorter clause-level units. Epistemically, Perplexity sits within the same broad search-grounded pattern of attenuated hedging relative to Vanilla GPT, but is less attenuated than the other two search systems: cognitive-process language remains high (\textit{cogproc}$=$0.082 vs.\ 0.087 Vanilla, 0.067 Search GPT, 0.044 AIO) and tentative language is likewise higher than the other search systems (\textit{tentat}$=$0.0218 vs.\ 0.025 Vanilla, 0.015 Search GPT, 0.010 AIO). Affect markers show the same search-grounded shift toward lower positive and comparatively higher negative emotion as the GPT-only systems (\textit{posemo}$=$0.034; \textit{negemo}$=$0.0108, the highest among the four systems).

\begin{table}[ht]
\small
\centering
\sffamily
\resizebox{\columnwidth}{!}{%
\begin{tabular}{lrrrr}
\rowcolor{blue!10}\textbf{Feature} & $\textbf{GPT}_\textbf{V}$ & $\textbf{GPT}_\textbf{S}$ & $\textbf{Google}_\textbf{AI}$ & \textbf{Perplexity} \\
\midrule
Word Count           & 87    & 140   & 83    & 190   \\
Avg.\ Sent.\ Length  & 16.3  & 18.0  & 11.4  & 14.2  \\
Coleman-Liau Index   & 11.5  & 12.8  & 11.7  & 12.8  \\
Type-Token Ratio     & 0.73  & 0.64  & 0.75  & 0.68  \\
\hdashline
cogproc               & 0.087 & 0.067 & 0.044 & 0.082 \\
tentat                & 0.025 & 0.015 & 0.010 & 0.022 \\
\hdashline
posemo                & 0.045 & 0.044 & 0.025 & 0.034 \\
negemo                & 0.008 & 0.009 & 0.006 & 0.011 \\
\bottomrule
\end{tabular}}
\caption{Descriptive comparison of Perplexity against the systems in \autoref{tab:rq2_main} on the subset of features computed for all four systems. Values are means; Perplexity is interpreted descriptively rather than via inferential tests.\vspace{-8pt}}
\label{tab:rq2_perplexity}
\end{table}

\begin{table*}[ht]
\small
\centering
\sffamily
\resizebox{0.8\textwidth}{!}{%
\begin{tabular}{@{}l@{}ccc}
\rowcolor{blue!10}& \textbf{Search GPT} & \textbf{Google AIO} & \textbf{Perplexity} \\
\rowcolor{blue!10}& \footnotesize Mean Diff (pp) [95\% CI] & \footnotesize Mean Diff (pp) [95\% CI] & \footnotesize Mean Diff (pp) [95\% CI] \\
\midrule
\rowcolor{gray!15}\multicolumn{4}{@{}l}{\textit{Document Length}} \\
\quad Short ($<$200 words) & \dvalci{-0.134}{-0.16}{-0.11}{^{***}} & \dvalci{-0.177}{-0.19}{-0.16}{^{***}} & \dvalci{-0.144}{-0.16}{-0.13}{^{***}} \\
\quad Medium (200--800) & \dvalci{+0.000}{-0.01}{+0.02}{} & \dvalci{+0.014}{+0.01}{+0.02}{^{**}} & \dvalci{+0.003}{-0.01}{+0.01}{} \\
\quad Long ($>$800 words) & \dvalci{+0.039}{+0.03}{+0.05}{^{***}} & \dvalci{+0.046}{+0.04}{+0.05}{^{***}} & \dvalci{+0.050}{+0.04}{+0.06}{^{***}} \\[2pt]
\rowcolor{gray!15}\multicolumn{4}{@{}l}{\textit{Source Sentiment}} \\
\quad Positive & \dvalci{+0.004}{-0.01}{+0.02}{} & \dvalci{+0.010}{+0.00}{+0.02}{^{**}} & \dvalci{+0.007}{+0.00}{+0.01}{^{*}} \\
\quad Neutral & \dvalci{-0.008}{-0.03}{+0.01}{} & \dvalci{-0.010}{-0.02}{+0.00}{} & \dvalci{-0.005}{-0.02}{+0.01}{} \\
\quad Negative & \dvalci{-0.037}{-0.10}{+0.03}{} & \dvalci{-0.138}{-0.17}{-0.11}{^{***}} & \dvalci{-0.101}{-0.13}{-0.07}{^{***}} \\[2pt]
\rowcolor{gray!15}\multicolumn{4}{@{}l}{\textit{Domain Category}} \\
\quad Encyclopedia & \dvalci{+0.022}{+0.01}{+0.03}{^{**}} & \dvalci{+0.050}{+0.04}{+0.06}{^{***}} & \dvalci{+0.088}{+0.08}{+0.10}{^{***}} \\
\quad Government & \dvalci{-0.042}{-0.08}{+0.00}{^{*}} & \dvalci{-0.034}{-0.06}{-0.01}{^{**}} & \dvalci{-0.046}{-0.06}{-0.03}{^{***}} \\
\quad Social/Forum & {--} & \dvalci{-0.221}{-0.25}{-0.20}{^{***}} & {--} \\
\quad Other & \dvalci{-0.015}{-0.03}{-0.00}{^{*}} & \dvalci{+0.001}{-0.01}{+0.01}{} & \dvalci{-0.019}{-0.03}{-0.01}{^{***}} \\[2pt]
\rowcolor{gray!15}\multicolumn{4}{@{}l}{\textit{Wiki vs.\ Non-Wiki}} \\
\quad Wikipedia & \dvalci{+0.026}{+0.01}{+0.04}{^{***}} & \dvalci{+0.054}{+0.04}{+0.07}{^{***}} & \dvalci{+0.096}{+0.08}{+0.11}{^{***}} \\
\quad Non-Wikipedia & \dvalci{-0.017}{-0.03}{-0.00}{^{**}} & \dvalci{-0.009}{-0.02}{-0.00}{^{**}} & \dvalci{-0.022}{-0.03}{-0.02}{^{***}} \\[2pt]
\rowcolor{gray!15}\multicolumn{4}{@{}l}{\textit{Citation Position}} \\
\quad Position 0 (first) & \dvalci{+0.015}{-0.00}{+0.03}{^{*}} & \dvalci{+0.032}{+0.02}{+0.05}{^{***}} & \dvalci{+0.128}{+0.11}{+0.14}{^{***}} \\
\quad Position 1 & \dvalci{+0.001}{-0.02}{+0.02}{} & \dvalci{+0.020}{+0.00}{+0.04}{^{*}} & \dvalci{+0.053}{+0.04}{+0.07}{^{***}} \\
\quad Position 2 & \dvalci{-0.021}{-0.04}{-0.00}{^{*}} & \dvalci{+0.012}{-0.01}{+0.03}{} & \dvalci{+0.010}{-0.00}{+0.02}{} \\[2pt]
\rowcolor{gray!15}\multicolumn{4}{@{}l}{\textit{Source Subjectivity}} \\
\quad High subjectivity & \dvalci{-0.023}{-0.05}{+0.00}{^{*}} & \dvalci{-0.032}{-0.05}{-0.02}{^{***}} & \dvalci{-0.038}{-0.05}{-0.02}{^{***}} \\
\quad Low subjectivity & \dvalci{+0.004}{-0.01}{+0.01}{} & \dvalci{+0.008}{+0.00}{+0.01}{^{**}} & \dvalci{+0.008}{+0.00}{+0.01}{^{**}} \\[2pt]
\bottomrule
\multicolumn{4}{@{}l}{\footnotesize ***\,$p<.001$, **\,$p<.01$, *\,$p<.05$. Brackets are 95\% percentile bootstrap CIs (n=1{,}000).}\\
\multicolumn{4}{@{}l}{\footnotesize Positive = over-represented; negative = under-represented.}\\
\end{tabular}}
\caption{Coverage Parity (CP) analysis across attribute schemes with 95\% bootstrap confidence intervals, extending \autoref{tab:rq3_cp} with Perplexity and interval estimates.\vspace{-8pt}}
\label{tab:rq3_cp_appendix}
\end{table*}

\begin{table}[ht]
\small
\centering
\sffamily
\resizebox{\columnwidth}{!}{%
\begin{tabular}{lcccccc}
& \multicolumn{2}{c}{\textbf{Search GPT}} & \multicolumn{2}{c}{\textbf{Google AIO}} & \multicolumn{2}{c}{\textbf{Perplexity}} \\
\cmidrule(lr){2-3} \cmidrule(lr){4-5} \cmidrule(lr){6-7}
\rowcolor{blue!10}\textbf{Topic} & EC & Coverage & EC & Coverage & EC & Coverage \\
\midrule
\rowcolor{gray!10}Business/Fin./Econ.    & 0.162 & 0.500 & 0.162 & 0.447 & 0.166 & 0.293 \\
Education/Lit.         & 0.189 & 0.481 & 0.175 & 0.445 & 0.169 & 0.316 \\
\rowcolor{gray!10}Entertainment          & 0.185 & 0.480 & 0.217 & 0.439 & 0.245 & 0.384 \\
History                & 0.160 & 0.460 & 0.169 & 0.414 & 0.151 & 0.315 \\
\rowcolor{gray!10}Music & 0.236 & 0.449 & 0.177 & 0.472 & 0.213 & 0.448 \\
Politics/News          & 0.152 & 0.506 & 0.168 & 0.438 & 0.169 & 0.313 \\
\rowcolor{gray!10}Religion               & 0.138 & 0.523 & 0.165 & 0.449 & 0.144 & 0.353 \\
Science                & 0.141 & 0.543 & 0.149 & 0.450 & 0.151 & 0.302 \\
\rowcolor{gray!10}Sports                 & 0.164 & 0.542 & 0.204 & 0.494 & 0.211 & 0.337 \\
Technology             & 0.131 & 0.535 & 0.189 & 0.436 & 0.175 & 0.315 \\
\rowcolor{gray!10}Travel/Geography       & 0.162 & 0.477 & 0.177 & 0.437 & 0.178 & 0.307 \\
\midrule
\textit{Overall}       & \textit{0.164} & \textit{0.500} & \textit{0.177} & \textit{0.447} & \textit{0.178} & \textit{0.332} \\
\bottomrule
\end{tabular}}
\caption{Equal Coverage (EC) and mean coverage probability (Coverage) by topic, extending \autoref{tab:rq3_ec} with Perplexity.\vspace{-8pt}}
\label{tab:rq3_ec_appendix}
\end{table}

\begin{table}[ht]
\small
\centering
\sffamily
\resizebox{0.8\columnwidth}{!}{
\begin{tabular}{lccccc}
\rowcolor{blue!10}\textbf{Topic} & \textbf{cogproc} & \textbf{tentat} & \textbf{posemo} & \textbf{formality} & \textbf{politeness} \\
\midrule
\rowcolor{gray!15}Technology          & \textbf{1.37} & 0.69           & \textbf{1.30} & \textbf{1.19} & \textbf{0.84} \\
Travel/Geography    & \textbf{0.88} & 0.64           & \textbf{1.01} & \textbf{1.76} & 0.79           \\
\rowcolor{gray!15}Sports              & 0.78           & 0.40           & \textbf{0.85} & \textbf{0.95} & \textbf{0.94} \\
Science             & 0.75           & 0.43           & \textbf{0.82} & \textbf{1.28} & 0.38           \\
\rowcolor{gray!15}Politics/News       & 0.56           & 0.37           & 0.76           & \textbf{1.21} & 0.02           \\
Music               & 0.40           & 0.17           & 0.62           & \textbf{1.26} & 0.34           \\
\rowcolor{gray!15}Religion            & 0.41           & 0.18           & 0.51           & \textbf{1.00} & 0.00           \\
Entertainment       & 0.28           & 0.16           & 0.32           & \textbf{0.82} & $-$0.06        \\
\rowcolor{gray!15}Education/Lit.      & 0.11           & $-$0.03        & 0.37           & \textbf{0.86} & 0.04           \\
History             & 0.20           & 0.07           & 0.29           & 0.76           & 0.09           \\
\rowcolor{gray!15}Business/Fin./Econ. & 0.12           & 0.03           & 0.39           & 0.68           & 0.11           \\
\midrule
\textit{Overall}    & \textit{0.48}  & \textit{0.27}  & \textit{0.61}  & \textit{0.99} & \textit{0.30} \\
\bottomrule
\end{tabular}}
\caption{Cohen's $d$ for Search GPT vs.\ Google AIO by topic. Positive $d$ indicates higher values in Search GPT. \textbf{Bold} := large effect ($|d|\!\geq\!0.8$). Topics are sorted by mean $|d|$ across the five features.\vspace{-8pt}}
\label{tab:rq2_topic_heatmap}
\end{table}

\para{RQ3: Source-summary coverage.} \autoref{tab:rq3_ec_appendix} and \autoref{tab:rq3_cp_appendix} extend \autoref{tab:rq3_ec} and \autoref{tab:rq3_cp} with Perplexity. Perplexity does not improve source balancing relative to the other systems: its EC is nearly identical to Google AIO and higher (worse, i.e.\ less even) than Search GPT ($0.178$ vs.\ $0.177$ AIO and $0.164$ Search GPT), while its overall coverage is the lowest of the three (0.332 vs.\ 0.447 AIO and 0.500 Search GPT). CP patterns match our main findings: short sources are strongly under-weighted and long sources over-weighted across all three systems; Wikipedia and encyclopedic sources are over-represented (Wikipedia: $+9.6$pp for Perplexity, the largest of the three systems); and government sources are under-represented in all three. Negative-sentiment sources are also under-covered in Perplexity ($-10.1$pp, $p<.001$), consistent with the pattern already observed for Google AIO. Because Perplexity's retained EC sample contains too few social/forum citations for a stable CP estimate (consistent with RQ1's near-zero social-source citation rate for Perplexity), we report the social/forum row only for Search GPT and Google AIO.

Overall, our findings reinforce that generative search systems differ systematically from vanilla LLM output and from traditional organic search along all three axes we study, and Perplexity reinforces the same directional findings while extending the external validity of our audit beyond OpenAI and Google.

\subsection{Prompts and Generation Settings}\label{app:prompts}

We include the main prompts used for response collection and ACU extraction to support reproducibility. For Vanilla GPT and Search GPT, we used GPT-4o-mini with temperature 0. Google AI Overview was collected through SerpAPI and is a black-box system for which we do not control the underlying prompt, model version, or decoding parameters.

\para{Vanilla GPT response collection:} The system message was:
\begin{quote}
\small
\texttt{You are a helpful assistant that provides accurate and concise answers to questions.}
\end{quote}
The user message was the raw search query.

\para{Search-enabled response collection:} For Search GPT and the Perplexity + Grok experiments, the system message differed from the Vanilla GPT prompt only in informing the model it had access to web search:
\begin{quote}
\small
\texttt{You are a helpful assistant that provides accurate and concise answers to questions. You have access to the web\_search tool for making web searches.}
\end{quote}
The user message was the raw search query. Web search was enabled through the corresponding search tool for each provider, and URL citation annotations returned by the APIs were retained for analysis.

\para{ACU extraction:} For ACU extraction, the system prompt was:
\begin{quote}
\small
\texttt{You are a helpful assistant that extracts atomic content units from text.}
\end{quote}
The user prompt used the following template:
\begin{quote}
\small
An atomic content unit (ACU) is a single, self-contained factual statement extracted from a summary. Each ACU should describe exactly one fact and not need further splitting. If a sentence contains compound subjects or objects, extract a separate ACU for each. Different ACUs may share overlapping content. Do not use reported speech. Number each ACU.

Below are examples.

Summary: The Great Wall of China was built over many centuries, with the most well-known sections constructed during the Ming Dynasty (1368--1644). It stretches approximately 13,171 miles and was primarily built to protect against invasions from northern tribes. The wall is a UNESCO World Heritage Site and one of the most visited tourist attractions in the world.

Extracted Atomic Content Units:
(1) The Great Wall of China was built over many centuries.
(2) The most well-known sections were constructed during the Ming Dynasty (1368--1644).
(3) It stretches approximately 13,171 miles.
(4) It was primarily built to protect against invasions from northern tribes.
(5) The wall is a UNESCO World Heritage Site.
(6) It is one of the most visited tourist attractions in the world.

Summary: Python and Java are both popular programming languages, but they differ in syntax, execution speed, and use cases. Python uses dynamic typing and is known for readability, while Java uses static typing and compiles to bytecode for the JVM. Python is widely used in data science and machine learning, whereas Java dominates enterprise applications and Android development.

Extracted Atomic Content Units:
(1) Python is a popular programming language.
(2) Java is a popular programming language.
(3) They differ in syntax.
(4) They differ in execution speed.
(5) They differ in use cases.
(6) Python uses dynamic typing.
(7) Python is known for readability.
(8) Java uses static typing.
(9) Java compiles to bytecode for the JVM.
(10) Python is widely used in data science.
(11) Python is widely used in machine learning.
(12) Java dominates enterprise applications.
(13) Java dominates Android development.

Now extract atomic content units for the following summary. Follow the same numbered format.

Summary: \{summary\}

Extracted Atomic Content Units:
\end{quote}

\section{Compute Resources}

All experiments on open-source models were run
on internal organization CPU and GPU servers equipped with 3xNVIDIA A40.

\end{document}